\documentclass[12pt]{amsart}
\usepackage{amssymb,latexsym}
\theoremstyle{plain}
\numberwithin{equation}{section}
\newtheorem{theorem}{Theorem}

\newtheorem{lemma}{Lemma}
\newtheorem{corollary}{Corollary}

\newtheorem{remark}{Remark}
\newtheorem{definition}{Definition}
\begin{document}

CFT-Meijo-Bedlevo-article.tex

\title[RHPWN and the Virasoro--Zamolodchikov--$w_{\infty}$ Algebra]{Renormalized Higher Powers of White Noise and the Virasoro--Zamolodchikov--$w_{\infty}$ Algebra}

\author{Luigi Accardi}
\address{Centro Vito Volterra, Universit\`{a} di Roma Torvergata\\
            via Columbia  2, 00133 Roma, Italy}
\thanks{The material in this paper was presented by Professor Luigi Accardi at  the``International Workshop on  White Noise Theory and its Applications''  held at  Meijo University in Nagoya, Japan, on September 18,  2006 and at the ``Workshop:  Non-Commutative Harmonic Analysis with applications to Probability" held in Bedlevo, Poland,  September 30 -- October 5, 2006  }
\email{accardi@volterra.mat.uniroma2.it}
\urladdr{http://volterra.mat.uniroma2.it}
\author{Andreas Boukas}
\address{Department of Mathematics and Natural Sciences, American College of Greece\\
 Aghia Paraskevi, Athens 15342, Greece}
\email{andreasboukas@acgmail.gr}

\date{\today}



\begin{abstract} Recently (cf. \cite{ABIDAQP06} and \cite{ABIJMCS06}) L. Accardi and A. Boukas proved that the generators of the second quantized  Virasoro--Zamolodchikov--$w_{\infty}$  algebra can be expressed in terms of the Renormalized Higher Powers of White Noise and conjectured that this inclusion might in fact be an identity, in the sense that the converse is also true. In this paper we prove that this conjecture is true.  We also explain the difference between this result and the Boson representation of the Virasoro algebra, which realizes, in the $1$--mode case (in particular without renormalization), an inclusion of this algebra into the full oscillator algebra.  This inclusion was known  in the physical literature and some heuristic results were obtained in the direction of  the extension of this inclusion to the $1$--mode Virasoro--Zamolodchikov--$w_{\infty}$ algebra. However the possibility of an identification  of the second quantizations of these two algebras was not even conjectured in the physics literature.  
\end{abstract}

\maketitle

\section{Introduction}

\begin{definition} The standard {\it $d$--dimensional Fock scalar White Noise}  is defined by a quadruple $\{  {\mathcal H} ,  b_t , b_t^+ , \Phi \}$, where $t\in\mathbb{R}^d$, ${\mathcal H}$ is a Hilbert space, $\Phi\in{\mathcal H}$ is a unit vector called the Fock vacuum, and $b_t$, $b^{\dagger}_t$ are operator valued Hida distributions satisfying  the Boson commutation relations

\[
[ b_t , b_s^{\dagger} ] = \delta(t-s) 
\] 

and having the Fock property

\[
 b_t \Phi = 0  
\]      

and the adjoint property

\[
 ( b_t^{\dagger})^{\dagger} = b_t      
\]
plus additional domain properties (not specified here).
\end{definition}

\begin{theorem} \label{exRHPWNliealg} Denoting ${\mathcal S}_0$ the space of  right--continuous step functions $f:\mathbb{R}\longrightarrow\mathbb{C}$ with compact support and satisfying $f(0)=0$, there exists a $*$--Lie algebra with generators given by:

\[
\{ B_k^n(f) \ : \ k,n\in\mathbb{N} \ , \ f\in {\mathcal S}_0 \} 
\]

involution given by:
 
\[
\left(B_k^n(f)\right)^{*} = B_n^k(\bar f)
\]

and brackets given by:

\[
[B^n_k(g),B^N_K(f) ]_{RHPWN}:= \left( k\,N- K\,n     \right)\, B^{n+N-1}_{k+K-1}( g f)
\]
\end{theorem}

\begin{proof} The  $*$--property is clear by construction. By direct calculations one shows that the brackets $[ \ \cdot \ , \ \cdot \ ]_{RHPWN}$ satisfy the Jacobi relations.  For details see \cite{ABIDAQP06} and  \cite{ABIJMCS06}.
\end{proof}

The $*$--Lie algebra defined by the above theorem is called the {\bf Renormalized Higher Powers of White Noise} (RHPWN) $*$--Lie algebra.  The following problems arise:

 (i) Construct a concrete mathematical model for the abstractly defined (RHPWN) $*$--Lie algebra.

 (ii) Construct Hilbert space representations for the (RHPWN) $*$--Lie algebra.

(iii) Prove exponentiability of the symmetric generators of the (RHPWN) $*$--Lie algebra in a given Hilbert space representation and identify the corresponding (Lie) group.

Heuristic results in the physics literature (cf. \cite{BK91}), and the results presented in this paper, suggest that a natural candidate for the corresponding (Lie) group is the group of area preserving diffeomorphisms on a (special) $2$--manifold (there are classical realizations on the cylinder $\mathbb{R}\times S^1 $).

The above given definition of  the RHPWN $*$--Lie algebra was motivated by the discovery that, if the powers of the Dirac delta function are renormalized by the prescription

\begin{eqnarray}
\delta^l(t-s)=\delta(s)\,\delta(t-s),\,\,\,\,\,l=2,3,....\label{delta}
\end{eqnarray}

then the resulting Renormalized Higher Powers of White Noise  

\[
B_k^n(f):=\int_{\mathbb{R}^d}\,f(t)\,{b_t^{\dagger}}^n\, b_t^k\,dt  \qquad ; \ k,n\in\mathbb{N}
\]
\[
B_0^0(f):=\int_{\mathbb{R}^d}\,f(t)\,dt\,\cdot\,1  \qquad \hbox{(multiple of the unique central element)}
\]

which are well defined as sesquilinear forms (matrix elements) on the algebraic span of the number vectors $\prod_{i=1}^{m}B^1_0(f_i)\,\Phi$ of the usual Boson Fock space, satisfy weakly on that domain, the defining relations of the  RHPWN $*$--Lie algebra in the following sense:

--  the adjoint $B_k^n(f)^* $ of the sesquilinear form $B_k^n(f)$ is defined in the obvious way

--  the brackets (commutator) $[B_k^n(f), B_K^N(g)]$ of the sesquilinear forms $B_k^n(f)$ and $B_K^N(g)$ is defined by bringing to normal order the products $B_k^n(f)\,B_K^N(g)$ and $B_K^N(g)\,B_k^n(f)$ by applying the RHPWN commutation relations, including the renormalization prescription, and letting the resulting normally ordered form act on the $1$--st order number vectors through the usual prescriptions.

The family of these sesquilinear forms is clearly a complex vector space with the pointwise operations and, with the above defined involution and brackets, it becomes a representation of the RHPWN $*$--Lie algebra introduced in Theorem \ref{exRHPWNliealg}.

\begin{definition}\label{top}  On the space of all sesquilinear forms on the algebraic span of the number vectors of the usual Boson Fock space, we define a topology by the semi--norms:

\begin{equation}
| q |_{x,y} := |q (x,y)|\label{sn}
\end{equation}

where $q$ is a sesquilinear form and $x,y$ are number vectors.
\end{definition}

\begin{lemma} \label{dfwnseries} Let $A_{n,k}$ be an arbitrary double sequence of complex numbers and let $f_{n,k}$ be an arbitrary double sequence in ${\mathcal S}_0$. Then the sesquilinear form 

\begin{equation}
q:=\sum_{n,k} A_{n,k}B_k^n(f_{n,k})\label{genrhpwn}
\end{equation}

is well defined (weakly) on the algebraic span of the number vectors of the usual  Boson Fock space.
\end{lemma}

\begin{proof} For any pair of number vectors $x,y$ 

\[
q (x,y) = \sum_{n,k} A_{n,k} \langle x, B_k^n(f_{n,k}) y \rangle
\]

is a finite sum of complex numbers.
\end{proof}

It follows from Lemma \ref{dfwnseries} that the completion of the RHPWN $*$--Lie algebra with respect to the family of seminorms (\ref{sn}) is the family of formal series of the form (\ref{genrhpwn}). In the following when speaking of the RHPWN $*$--Lie algebra we will mean this larger family. Lemma \ref{dfwnseries} allows one to give a meaning to a rather general class of functions of the renormalized white noise.
 
\begin{definition}\label{basic} Let $F(b_t^{\dagger} , b_t)$ be a formal power series in the noncommutative indeterminates $b_t^{\dagger} , b_t$. If, by applying the RHPWN commutation relations, including the renormalization prescriptions, one can write this series in the form

\[
\sum_{n,k} A_{n,k} b_t^{{\dagger} n}b_t^{k}
\]

where each coefficient $A_{n,k}$ is a complex number (in particular finite!), then we say that the formal power series $F(b_t^{\dagger} , b_t)$ defines {\bf a function of $b_t^{\dagger} , b_t$}. The meaning of this function is that by multiplying by test functions $f_{n,k}(t)$ and integrating term by term in $dt$, one obtains the sesquilinear form $\sum_{n,k} A_{n,k}B_k^n(f_{n,k}) $, which is well defined by Lemma \ref{dfwnseries}.

\end{definition}

In the following we will produce concrete examples of  functions of $b_t^{\dagger} , b_t$.

\section{The Virasoro--Zamolodchikov--$w_{\infty}$  $*$--Lie algebra } 

Following a completely different line of thought, people in conformal field theory and in string theory were led to introduce another $*$--Lie algebra (cf.  \cite{BK91}, \cite{K95}).

\begin{theorem}\label{T-1} There exists a $*$--Lie algebra with generators 

\[
\{ \hat B_k^n \ : \ n\in\mathbb{N} \ , \  n\geq2 \ , \ k\in\mathbb{Z} \} 
\]

involution given by:
 
\[
\left(\hat{B}_k^n\right)^{*} = \hat{B}_{-k}^n
\]

and brackets given by:

\[
[\hat{B}^n_k,\hat{B}^N_K ]_{ w_{\infty}} =
 \left( k\,(N-1)- K\,(n-1) \right)\, \hat{B}^{n+N-2}_{k+K}
\]

\end{theorem} 
\begin{proof} See \cite{K95}.
\end{proof}

\begin{definition} The $*$--Lie algebra defined in Theorem \ref{T-1} is called  the {\bf Virasoro--Zamolodchikov--$w_{\infty}$ } $*$--Lie algebra.
\end{definition}

Notice that no test functions appear in the above definition.  In our language, we can say that the above $*$--Lie algebra is the $1$--mode version of the algebra we are interested in, or equivalently, that the algebra we are interested in, is a second quantization of the 
$w_{\infty}$ $*$--Lie algebra. As usual, the existence of such an object has to be proved.

\begin{theorem}\label{T0} There exists a $*$--Lie algebra with generators 

\[
 \{ \hat B_k^n(f) \ : \ n\in\mathbb{N} \ , \  n\geq2 \ , \ k\in\mathbb{Z} \ , \ f\in {\mathcal S}_0 \} 
\]

involution given by:
 
\begin{equation}
\left(\hat{B}_k^n(f)\right)^{*} = \hat{B}_{-k}^n(\bar f)\label{inv}
\end{equation}

and brackets given by:

\begin{equation}
[\hat{B}^n_k( g),\hat{B}^N_K(f) ]_{ w_{\infty}} =
 \left( k\,(N-1)- K\,(n-1) \right)\, \hat{B}^{n+N-2}_{k+K}( g f)\label{comm}
\end{equation}
\end{theorem}

\begin{proof} 

Clearly, for all test functions $f,g\in {\mathcal S}_0$ and $n,k, N, K\geq0$, 

\[
[\hat B^N_K( g),\hat B^N_K(f) ]_{w_{\infty}}=0
\]

and

\[
[\hat B^N_K( g),\hat B^n_k(f) ]_{w_{\infty}}=-[\hat B^n_k(f),\hat B^N_K( g) ]_{w_{\infty}}
\]

To show that commutation relations (\ref{comm}) satisfy the Jacobi identity we must show that  for all test functions $f,g,h$ and $n_i,k_i\geq0$, where  $i=1,2,3$,

\[
\left[\hat B^{n_1}_{k_1}(f),[\hat B^{n_2}_{k_2}(g),\hat B^{n_3}_{k_3}(h)]_{w_{\infty}}\right]_{w_{\infty}}+ \left[\hat B^{n_3}_{k_3}(h),[\hat B^{n_1}_{k_1}(f),\hat B^{n_2}_{k_2}(g)]_{w_{\infty}}\right]_{w_{\infty}}
\]

\[
+\left[\hat B^{n_2}_{k_2}(g),[\hat B^{n_3}_{k_3}(h),\hat B^{n_1}_{k_1}(f)]_{w_{\infty}}\right]_{w_{\infty}}  =0
\]

i.e.  that

\[
\{(k_2(n_3-1)-k_3(n_2-1))(k_1(n_2+n_3-3)-(k_2+k_3)(n_1-1))+
\]

\[
 (k_1(n_2-1)-k_2(n_1-1))(k_3(n_1+n_2-3)-(k_1+k_2)(n_3-1))+
\]

\[
(k_3(n_1-1)-k_1(n_3-1))(k_2(n_3+n_1-3)-(k_3+k_1)(n_2-1))\}\,\hat B^{n_1+n_2+n_3-4}_{k_1+k_2+k_3}(fgh)=0
\]

which is  true, since

\[
(k_2(n_3-1)-k_3(n_2-1))(k_1(n_2+n_3-3)-(k_2+k_3)(n_1-1))+
\]

\[
 (k_1(n_2-1)-k_2(n_1-1))(k_3(n_1+n_2-3)-(k_1+k_2)(n_3-1))+
\]

\[
(k_3(n_1-1)-k_1(n_3-1))(k_2(n_3+n_1-3)-(k_3+k_1)(n_2-1))=0
\]

Finally, in order to show that, with involution defined by (\ref{inv}),

\[
[\hat B^n_k(f),\hat B^N_K(g)]_{w_{\infty}}^*=[\left(\hat B^N_K(g)\right)^*,\left(\hat B^n_k(f)\right)^* ]_{w_{\infty}}
\]

i.e that

\[
[\hat B^n_k(f),\hat B^N_K(g)]_{w_{\infty}}^*=[\hat B^N_{-K}(\bar g),\hat B^n_{-k}(\bar f) ]_{w_{\infty}}
\]

we notice that both sides of the above equation are equal to

\[
(k(N-1)-K(n-1))\,\hat B^{n+N-2}_{-(k+K)}(\overline{fg})
\]

\end{proof}

\begin{definition}  The $*$--Lie algebra defined in Theorem \ref{T0} is called the {\bf second quantized  Virasoro--Zamolodchikov-- $w_{\infty}$} $*$--Lie algebra.
\end{definition}

The term Zamolodchikov is due to the fact that $w_{\infty}$ is a large $N$ limit of Zamolodchikov's $W_N$ algebra (cf \cite{K95} and \cite{Z85}). The term Virasoro is justified by the following theorem.

\begin{theorem} The family of operators

\[
 \{ \hat{B}^2_k \ :  \ k\in\mathbb{Z} \}  
\]

forms a $*$--Lie sub--algebra of the $w_{\infty}$  Lie algebra with  involution

\[
\left(\hat{B}_k^2\right)^{*} = \hat{B}_{-k}^2
\]

and brackets

\[
[\hat{B}^2_k,\hat{B}^2_{K} ]_{Vir}:=(k-K)\,\hat{B}^{2}_{k+K}
\]
\end{theorem}

\begin{remark} These are precisely the defining relations of the centerless Virasoro (or Witt) algebra.\end{remark} 

\begin{proof} The proof follows directly from Theorem \ref{T-1} for $n=N=2$.

\end{proof}

The following second quantized version of the above theorem also holds.

\begin{theorem}\label{sqvir}  The family of operators

\[
\{ \hat{B}^2_k( f) \ : \ f\in {\mathcal S}_0 \ ; \ k\in\mathbb{Z} \}  
\]

forms a $*$--Lie sub--algebra of the second quantized $w_{\infty}$ Lie algebra with  involution

\[
\left(\hat{B}_k^2(f)\right)^{*} = \hat{B}_{-k}^2(\bar f)
\]

and brackets

\[
[\hat{B}^2_k(g),\hat{B}^2_{K}(f) ]_{Vir}:=(k-K)\,\hat{B}^{2}_{k+K}(g f)
\]

\end{theorem}

\begin{proof}
The proof follows directly from Theorem \ref{T0} for $n=N=2$.
\end{proof}

\begin{definition}  The $*$--Lie algebra defined in Theorem \ref{sqvir} is called the  second quantized centerless Virasoro (or Witt) algebra.
\end{definition}

\section{The connection between the second quantized $w_{\infty}$ and the RHPWN $*$--Lie algebras  }

The striking similarity between the brackets of the RHPWN $*$--Lie algebra 

\[
[B^n_k(g),B^N_K(f) ]_{RHPWN}:= \left( k\,N- K\,n     \right)\, B^{n+N-1}_{k+K-1}( g f)
\]

and the brackets of the second quantized $w_{\infty}$ $*$--Lie algebra 

\[
[\hat{B}^n_k( g),\hat{B}^N_K(f) ]_{ w_{\infty}} =
 \left( k\,(N-1)- K\,(n-1) \right)\, \hat{B}^{n+N-2}_{k+K}( g f)
\]

strongly suggests that there should be a connection between the two. However there are also strong dissimilarities. The sets, indexing the generators of the two algebras, are different:

\[
\{ B_k^n(f) \ : \ k,n\in\mathbb{N} \ , \ f\in {\mathcal S}_0 \} 
\]

for the RHPWN $*$--Lie algebra, and

\[
 \{ \hat B_k^n(f) \ : \ n\in\mathbb{N} \ , \  n\geq2 \ , \ k\in\mathbb{Z} \ , \ f\in {\mathcal S}_0 \} 
\]

for the second quantized $w_{\infty}$ $*$--Lie algebra . The involutions in the two algebras also look quite different:

\[
\left(\hat{B}_k^n(f)\right)^{*} = \hat{B}_{n}^k(\bar f)
\]

for the RHPWN $*$--Lie algebra, and

\[
\left(\hat{B}_k^n(f)\right)^{*} = \hat{B}_{-k}^n(\bar f)
\]

for the second quantized $w_{\infty}$ $*$--Lie algebra.  The following Theorem \ref{original},  obtained in \cite{ABIDAQP06}, was the first definite result in the direction of establishing a connection between the RHPWN and $w_{\infty}$ $*$--Lie algebras. 

\begin{theorem}\label{original}  In the sense of formal power series the following identity holds:

\begin{equation}\label{bhatfctnb}
\hat{B}_k^n(f)=
\int_{\mathbb{R}^d}\,f(t)\,e^{ \frac{k}{2}(b_t- b_t^{\dagger})}
\left(\frac{ b_t+ b_t^{\dagger}}{2}\right)^{n-1} \,  e^{ \frac{k}{2}(b_t- b_t^{\dagger})}\,dt
\end{equation}

\end{theorem}

\begin{proof} The proof can be found in \cite{ABIDAQP06}.
\end{proof}

The  following Lemma (cf \cite{ABIJMCS06}) on the generators of the Heisenberg--Weyl Lie algebra, is the basic tool used in the proofs of Theorems  \ref{original} above and \ref{T}, \ref{T2} below. 

\begin{lemma}\label{L} 
Let $x$ , $D$  and $h$  satisfy the Heisenberg commutation relations

\[
[D,x]=h, \,\,\,[D,h]=[x,h]=0
\]

Then, for all $s,a,c\in\mathbb{C}$

\[
e^{s(x+aD+ch)}=e^{sx}e^{saD}e^{(sc+\frac{s^2a}{2})h}
\]

and

\[
e^{sD}e^{ax}=e^{ax}e^{sD}e^{ash}
\]

\end{lemma}

\subsection{The inclusion: analytic continuation of the {\it second quantized} $w_{\infty}$ $ \subseteq $ RHPWN  }  The following Theorem expresses the generators $\hat{B}_k^n(f)$ of the second quantized $w_{\infty}$ $*$--Lie algebra as a series of the form $\sum_{n,k} A_{n,k}B_k^n(f_{n,k})$ of the generators $B_k^n(f)$ of the RHPWN $*$--Lie algebra. The considerations of the previous section show that this series has a meaning and that it is obtained from  a  function of $b_t$ and $b_t^{\dagger}$, also defined in that section. 

\begin{theorem}\label{T} 
Let $n\geq2$ and $k\in\mathbb{Z}$. Then, for all $f\in{\mathcal S}_0$,

\begin{equation}
\hat{B}^n_k( f)=\frac{1}{2^{n-1}}\,\sum_{m=0}^{n-1}\,
\binom{n-1}{m}
\, \sum_{p=0}^{\infty}\, \sum_{q=0}^{\infty}\, (-1)^p\,\frac{k^{p+q}}{p!\,q!}\,B^{ m+p}_{n-1-m+q}(f)\label{BhatfctB}
\end{equation}

where convergence of infinite sums is understood in 
the topology introduced in Definition \ref{top}, and the case $k=0$ is interpreted as

\[
\hat{B}^n_0( f)=\frac{1}{2^{n-1}}\,\sum_{m=0}^{n-1}\,
\binom{n-1}{m}
 \,B^{ m}_{n-1-m}(f)   
\]

\end{theorem}

\begin{proof} 
For fixed $t,s\in\mathbb{R}$,  we will make repeated use of Lemma \ref{L} with $D= b_t$, $x=  b_s^{\dagger}$ and $h=\delta(t-s)$.  We have 

\[
\hat{B}^n_k( f)=\int_{\mathbb{R}^d}\,f(s)\,e^{ \frac{k}{2}(b_s- b_s^{\dagger})}\left(\frac{ b_s+ b_s^{\dagger}}{2}\right)^{n-1} \,  e^{ \frac{k}{2}(b_s- b_s^{\dagger})}\,ds
\]

\[
=\frac{1}{2^{n-1}}\,\int_{\mathbb{R}^d}\,\int_{\mathbb{R}^d}\,f(t)\,e^{ \frac{k}{2}(b_t- b_s^{\dagger})}\left( b_t+ b_s^{\dagger}\right)^{n-1} \,  e^{ \frac{k}{2}(b_t- b_s^{\dagger})}\,\delta(t-s)\,dt\,ds
\]

\[
=\frac{1}{2^{n-1}}\,\int_{\mathbb{R}^d}\,\int_{\mathbb{R}^d}\,f(t)\,e^{ -\frac{k}{2}(b_s^{\dagger}-b_t)}\left( \frac{ \partial^{n-1} }{ \partial w^{n-1}}|_{w=0}e^{w (b_t+ b_s^{\dagger})}\right)\,  e^{ -\frac{k}{2}( b_s^{\dagger}-b_t)}\,\delta(t-s)\,dt\,ds
\]

\[
=\frac{1}{2^{n-1}}\,\frac{ \partial^{n-1} }{ \partial w^{n-1}}|_{w=0}\,\int_{\mathbb{R}^d}\,\int_{\mathbb{R}^d}\,f(t)\,e^{ -\frac{k}{2}(b_s^{\dagger}-b_t)}e^{w (b_t+ b_s^{\dagger})}\,  e^{ -\frac{k}{2}( b_s^{\dagger}-b_t)}\,\delta(t-s)\,dt\,ds
\]

\[
=\frac{1}{2^{n-1}}\,\frac{ \partial^{n-1} }{ \partial w^{n-1}}|_{w=0}\,\int_{\mathbb{R}^d}\,\int_{\mathbb{R}^d}\,f(t)\,e^{ -\frac{k}{2}\,b_s^{\dagger}}\,e^{ \frac{k}{2}b_t}\,e^{w \,b_s^{\dagger}}\, e^{w \,b_t}\, e^{ -\frac{k}{2}\, b_s^{\dagger}}\, e^{ \frac{k}{2}\, b_t}\,  e^{ \left(\frac{w^2}{2}-\frac{k^2}{4}\right)\,\delta(t-s)}\ \,\delta(t-s)\,dt\,ds
\]

\[
=\frac{1}{2^{n-1}}\,\frac{ \partial^{n-1} }{ \partial w^{n-1}}|_{w=0}\,\int_{\mathbb{R}^d}\,\int_{\mathbb{R}^d}\,f(t)\,e^{ -\frac{k}{2}\,b_s^{\dagger}}\,e^{w \,b_s^{\dagger}}\,e^{ \frac{k}{2}b_t} \, e^{ -\frac{k}{2}\, b_s^{\dagger}}\,e^{w \,b_t}\, e^{ \frac{k}{2}\, b_t}\,  e^{ \left(\frac{w^2}{2}-\frac{k^2}{4}\right)\,\delta(t-s)}\ \,\delta(t-s)\,dt\,ds
\]

\[
=\frac{1}{2^{n-1}}\,\frac{ \partial^{n-1} }{ \partial w^{n-1}}|_{w=0}\,\int_{\mathbb{R}^d}\,\int_{\mathbb{R}^d}\,f(t)\,e^{ \left(w-\frac{k}{2}\right)\,b_s^{\dagger}}\,e^{ \frac{k}{2}b_t} \, e^{ -\frac{k}{2}\, b_s^{\dagger}}\,e^{\left(w+\frac{k}{2}\right) \,b_t}\, e^{ \left(\frac{w^2}{2}-\frac{k^2}{4}\right)\,\delta(t-s)}\ \,\delta(t-s)\,dt\,ds
\]

\[
=\frac{1}{2^{n-1}}\,\frac{ \partial^{n-1} }{ \partial w^{n-1}}|_{w=0}\,\int_{\mathbb{R}^d}\,\int_{\mathbb{R}^d}\,f(t)\,e^{ \left(w-\frac{k}{2}\right)\,b_s^{\dagger}}\,e^{ -\frac{k}{2}\, b_s^{\dagger}}\, e^{ \frac{k}{2}b_t} \,e^{\left(w+\frac{k}{2}\right) \,b_t}\, e^{ \left(\frac{w^2}{2}-\frac{k^2}{2}\right)\,\delta(t-s)}\ \,\delta(t-s)\,dt\,ds
\]

\[
=\frac{1}{2^{n-1}}\,\frac{ \partial^{n-1} }{ \partial w^{n-1}}|_{w=0}\,\int_{\mathbb{R}^d}\,\int_{\mathbb{R}^d}\,f(t)\,e^{ \left(w-k\right)\,b_s^{\dagger}} \,e^{\left(w+k\right) \,b_t}\, e^{ \left(\frac{w^2}{2}-\frac{k^2}{2}\right)\,\delta(t-s)}\ \,\delta(t-s)\,dt\,ds
\]

\[
=\frac{1}{2^{n-1}}\,\frac{ \partial^{n-1} }{ \partial w^{n-1}}|_{w=0}\,\int_{\mathbb{R}^d}\,\int_{\mathbb{R}^d}\,f(t)\,e^{ \left(w-k\right)\,b_s^{\dagger}} \,e^{\left(w+k\right) \,b_t}\,  \sum_{m=0}^{\infty}\frac{  \left(\frac{w^2}{2}-\frac{k^2}{2}\right)^m}{m!}\,\delta^m (t-s)   \,\delta(t-s)\,dt\,ds
\]

\[
=\frac{1}{2^{n-1}}\,\frac{ \partial^{n-1} }{ \partial w^{n-1}}|_{w=0}\,\int_{\mathbb{R}^d}\,\int_{\mathbb{R}^d}\,f(t)\,e^{ \left(w-k\right)\,b_s^{\dagger}} \,e^{\left(w+k\right) \,b_t}\,   \sum_{m=0}^{\infty}\frac{  \left(\frac{w^2}{2}-\frac{k^2}{2}\right)^m}{m!}\,\delta^{m+1}(t-s)  \,dt\,ds
\]

\[
=\frac{1}{2^{n-1}}\,\frac{ \partial^{n-1} }{ \partial w^{n-1}}|_{w=0}\,\int_{\mathbb{R}^d}\,\int_{\mathbb{R}^d}\,f(t)\,e^{ \left(w-k\right)\,b_s^{\dagger}} \,e^{\left(w+k\right) \,b_t}\,\left(\delta (t-s)+    \sum_{m=1}^{\infty}\frac{  \left(\frac{w^2}{2}-\frac{k^2}{2}\right)^m}{m!}\,\delta(s)\,\delta(t-s)      \right)  \,dt\,ds
\]

\[
=\frac{1}{2^{n-1}}\,\frac{ \partial^{n-1} }{ \partial w^{n-1}}|_{w=0}\,\int_{\mathbb{R}^d}\,\int_{\mathbb{R}^d}\,f(t)\,e^{ \left(w-k\right)\,b_s^{\dagger}} \,e^{\left(w+k\right) \,b_t}\,\delta(t-s)\,dt\,ds
\]

since, by the assumption $f(0)=0$,  all $\delta(s)\,\delta(t-s)$ terms vanish.  Thus, using Leibniz's rule,

\[
\hat{B}^n_k( f)=\frac{1}{2^{n-1}}\,\frac{ \partial^{n-1} }{ \partial w^{n-1}}|_{w=0}\,\int_{\mathbb{R}^d}\,f(s)\,e^{ \left(w-k\right)\,b_s^{\dagger}} \,e^{\left(w+k\right) \,b_s}\,ds
\]

\[
=\frac{1}{2^{n-1}}\,\sum_{m=0}^{n-1}\,\binom{n-1}{m}
 \,\int_{\mathbb{R}^d}\,f(s)\,\frac{ \partial^{m} }{ \partial w^{m}}|_{w=0}\left(e^{ \left(w-k\right)\,b_s^{\dagger}} \right)\, \frac{ \partial^{n-1-m} }{ \partial w^{n-1-m}}|_{w=0}\left(  e^{\left(w+k\right) \,b_s}\right)\,ds
\]

\[
=\frac{1}{2^{n-1}}\,\sum_{m=0}^{n-1}\,
\binom{n-1}{m}
 \,\int_{\mathbb{R}^d}\,f(s)\, {b_s^{\dagger}}^m  \, e^{ -k\,b_s^{\dagger}} \, {b_s}^{n-1-m}\, e^{k\,b_s}\,ds
\]

\[
=\frac{1}{2^{n-1}}\,\sum_{m=0}^{n-1}\,
\binom{n-1}{m}
 \,\int_{\mathbb{R}^d}\,f(s)\,  {b_s^{\dagger}}^m  \,   \sum_{p=0}^{\infty}\, \frac{(-k)^p}{p!}\,{b_s^{\dagger}}^p \,  {b_s}^{n-1-m}   \,\sum_{q=0}^{\infty}\, \frac{k^q}{q!}\,{b_s}^q        \,ds
\]

\[
=\frac{1}{2^{n-1}}\,\sum_{m=0}^{n-1}\,
 \binom{n-1}{m}
 \, \sum_{p=0}^{\infty}\, \sum_{q=0}^{\infty}\, (-1)^p\,\frac{k^{p+q}}{p!\,q!}\,
\int_{\mathbb{R}^d}\,f(s)\,  {b_s^{\dagger}}^{m+p} \,  {b_s}^{n-1-m+q} \,ds
\]

\[
=\frac{1}{2^{n-1}}\,\sum_{m=0}^{n-1}\,
\binom{n-1}{m}
\, \sum_{p=0}^{\infty}\, \sum_{q=0}^{\infty}\, (-1)^p\,\frac{k^{p+q}}{p!\,q!}\,B^{ m+p}_{n-1-m+q}(f)
\]

\end{proof}

From the identity (\ref{BhatfctB}), it is clear that one can analytically continue the parameter $k$, in the definition of $\hat{B}_k^n(f)$, to an arbitrary complex number $k\in\mathbb{C}$ and $n\geq1$. After this extension the identity 

\[
[\hat{B}^1_k( g),\hat{B}^1_K(f) ]_{ w_{\infty} }=0
\]

still holds. Moreover, the proof of Theorem \ref{T} immediately extends to $k\in\mathbb{C}$ and we have:

\begin{theorem}\label{TT} 
Let $n\geq2$ and $z\in\mathbb{C}$. Then, for all $f\in{\mathcal S}_0$,

\begin{equation}
\hat{B}^n_z( f)=\frac{1}{2^{n-1}}\,\sum_{m=0}^{n-1}\,
\binom{n-1}{m}
\, \sum_{p=0}^{\infty}\, \sum_{q=0}^{\infty}\, (-1)^p\,\frac{z^{p+q}}{p!\,q!}\,B^{ m+p}_{n-1-m+q}(f)\label{BhatfctB2}
\end{equation}

where convergence of infinite sums is understood in 
the topology introduced in Definition \ref{top}, and the case $z=0$ is interpreted as

\[
\hat{B}^n_0( f)=\frac{1}{2^{n-1}}\,\sum_{m=0}^{n-1}\,
\binom{n-1}{m}
 \,B^{ m}_{n-1-m}(f)   
\]

\end{theorem}

\begin{corollary}\label{d} For $z\in\mathbb{C}$ and $k\in \{0,1,...\}$, 

\begin{equation}\label{der1}
\frac{ \partial^{k}}{\partial {z}^{k}}|_{z =0}\hat{B}^n_z(f):=\sum_{m=0}^k \, 
\frac{\binom{k}{m}}{2^{k+n-1} }\int_{\mathbb{R}^d}\,f(t)\,{\left(b_t-b_t^{\dagger}\right)}^m\,  \, 
{\left(b_t+b_t^{\dagger}\right)}^{n-1}\,{ \left(b_t-b_t^{\dagger}\right)}^{k-m}\,dt
\end{equation}

Moreover, for $k\geq1$

\begin{equation}\label{der}
\frac{ \partial^{k}}{\partial {z}^{k}}|_{z =0}\hat{B}^n_z(f)=\frac{1}{2^{n-1}}\,\sum_{m=0}^{n-1}\,
\binom{n-1}{m}
\, \sum_{p=0}^{\infty}\,\, (-1)^p\,\binom{k}{p}\,B^{ m+p}_{n+k-m-1-p}(f)
\end{equation}

\end{corollary}

\begin{proof} (\ref{der1}) is obtained from  (\ref{bhatfctnb})  with the use of Leibniz's rule. Finally, (\ref{der}) follows from (\ref{BhatfctB2}), by differentiating term-by-term $k$-times with respect to $z$, and noticing that only the terms $p+q=k$ will contribute.

\end{proof}

\subsection{The inclusion: RHPWN  $ \subseteq $ analytic continuation of the {\it second quantized} $w_{\infty}$} In this subsection we will find the expression of the generators $B^n_k( f)$ of the RHPWN $*$--Lie--algebra in terms of the generators $\hat{B}^n_z( f)$ of the analytically continued $w_{\infty}$$*$--Lie--algebra  , thus completing the identification of the two algebras.

\begin{theorem}\label{T2} Let $n,k\in\mathbb{N}\cup 0$. 
Then, for all $f\in{\mathcal S}_0$,

\begin{equation}\label{winfrhpwn}
B^n_k( f)=\sum_{\rho=0}^k \,\sum_{\sigma=0}^n 
\binom{k}{\rho}\binom{n}{\sigma}
 \frac{(-1)^{\rho}}{2^{\rho+\sigma}}\,
\frac{\partial^{\rho+\sigma }}{\partial {z}^{\rho+\sigma}}|_{z =0} \,\hat{B}^{k+n+1-(\rho+\sigma)}_{z}(f)
\end{equation}

where, as in Theorem \ref{T},  convergence of infinite sums is understood in the topology introduced in Definition \ref{top}. Moreover, the right hand side of (\ref{winfrhpwn}) is well defined in the sense of Definition 
 \ref{basic}.
\end{theorem}

\begin{proof} 
For $t,s\geq 0$, let $p_{t,s}:=b_t-b_s^{\dagger}$ and $q_{t,s}:=b_t+b_s^{\dagger}$. Then

\[
B_k^n(f)=\int_{\mathbb{R}^d}\,f(t)\,{b_t^{\dagger}}^n\, b_t^k\,dt
\]

\[
=\frac{1}{2^{n+k}}\,\int_{\mathbb{R}^d}\,\int_{\mathbb{R}^d}\,f(s)\,\left( p_{t,s}+q_{t,s} \right)^n\,\left(q_{t,s}- p_{t,s} \right) ^k\,\delta(s-t)\,ds\,dt
\]

\[
=\frac{1}{2^{n+k}}\,\int_{\mathbb{R}^d}\,\int_{\mathbb{R}^d}\,f(s)\,\frac{ \partial^{n} }{ \partial {\lambda}^{n}}|_{\lambda =0}\left(e^{ \lambda\,( p_{t,s}+q_{t,s})}\right)\, \frac{ \partial^{k} }{ \partial {\mu}^{k}}|_{\mu =0}\left(e^{ \mu \,( q_{t,s}- p_{t,s} )}\right)\,\delta(s-t)\,ds\,dt
\]

\[
=\frac{1}{2^{n+k}}\, \frac{ \partial^{n+k} }{ \partial {\lambda}^{n}\,{\mu}^{k}}|_{\lambda =\mu=0}    \,\int_{\mathbb{R}^d}\,\int_{\mathbb{R}^d}\,f(s)\,e^{ \lambda\,( p_{t,s}+q_{t,s})}\, e^{ \mu \,( q_{t,s}- p_{t,s} )}\,\delta(s-t)\,ds\,dt
\]

which, using Lemma \ref{L} with $D=p_{t,s} $, $x= q_{t,s}$ and $h=2\,\delta(s-t)  $, is equal to

\[
=\frac{1}{2^{n+k}}\, \frac{ \partial^{n+k} }{ \partial {\lambda}^{n}\,{\mu}^{k}}|_{\lambda =\mu=0}    \,\int_{\mathbb{R}^d}\,\int_{\mathbb{R}^d}\,f(s)\,e^{ \lambda\, p_{t,s}}\, e^{ \lambda\, q_{t,s}}\, e^{ \mu \, q_{t,s}}\,e^{ -\mu \, p_{t,s} }\,e^{ (\lambda^2-\mu^2)\,\delta(s-t)}  \,\delta(s-t)\,ds\,dt
\]

\[
=\frac{1}{2^{n+k}}\, \frac{ \partial^{n+k} }{ \partial {\lambda}^{n}\,{\mu}^{k}}|_{\lambda =\mu=0}    \,\int_{\mathbb{R}^d}\,\int_{\mathbb{R}^d}\,f(s)\,e^{ \lambda\, p_{t,s}}\, e^{( \lambda +\mu)\, q_{t,s}}\,e^{ -\mu \, p_{t,s} }\,e^{ (\lambda^2-\mu^2)\,\delta(s-t)}  \,\delta(s-t)\,ds\,dt
\]

which, again by Lemma \ref{L}, is equal to

\[
=\frac{1}{2^{n+k}}\, \frac{ \partial^{n+k} }{ \partial {\lambda}^{n}\,{\mu}^{k}}|_{\lambda =\mu=0}    \,\int_{\mathbb{R}^d}\,\int_{\mathbb{R}^d}\,f(s)\,e^{ \lambda\, p_{t,s}}\,e^{ -\mu \, p_{t,s} }\, e^{( \lambda +\mu)\, q_{t,s}}\,e^{ (\lambda+\mu)^2\,\delta(s-t)}  \,\delta(s-t)\,ds\,dt
\]

\[
=\frac{1}{2^{n+k}}\, \frac{ \partial^{n+k} }{ \partial {\lambda}^{n}\,{\mu}^{k}}|_{\lambda =\mu=0}    \,\int_{\mathbb{R}^d}\,\int_{\mathbb{R}^d}\,f(s)\,e^{ (\lambda-\mu)\, p_{t,s}}\, e^{( \lambda +\mu)\, q_{t,s}}\,e^{ (\lambda+\mu)^2\,\delta(s-t)}  \,\delta(s-t)\,ds\,dt
\]

and since, as in the proof of Theorem \ref{T}, only the constant term $1$ in the exponential series $e^{ (\lambda+\mu)^2\,\delta(s-t)}  $ will eventually make a nonzero contribution, the above is equal to

\[
=\frac{1}{2^{n+k}}\, \frac{ \partial^{n+k} }{ \partial {\lambda}^{n}\,{\mu}^{k}}|_{\lambda =\mu=0}    \,\int_{\mathbb{R}^d}\,\int_{\mathbb{R}^d}\,f(s)\,e^{ (\lambda-\mu )\, p_{t,s}}\, e^{( \lambda +\mu)\, q_{t,s}}\,\delta(s-t)\,ds\,dt
\]

which, using Leibniz's rule first for the derivative with respect to $\mu$ and then for the derivative with respect to $\lambda$, is equal to

\[
=\frac{1}{2^{n+k}}\,\sum_{\rho=0}^k 
\binom{k}{\rho}
 (-1)^{\rho} \,\frac{ \partial^{n} }{ \partial {\lambda}^{n}}|_{\lambda =0}    
\,\int_{\mathbb{R}^d}\,\int_{\mathbb{R}^d}\,f(s)\,p_{t,s}^{\rho}\,
e^{ \lambda\, p_{t,s}}\,q_{t,s}^{k-\rho}\, e^{ \lambda \, q_{t,s}}\,\delta(s-t)\,ds\,dt
\]

\[
=\frac{1}{2^{n+k}}\,\sum_{\rho=0}^k \,\sum_{\sigma=0}^n 
\binom{k}{\rho}\binom{n}{\sigma}
 (-1)^{\rho} \,\int_{\mathbb{R}^d}\,
\int_{\mathbb{R}^d}\,f(s)\,p_{t,s}^{\rho}\,p_{t,s}^{\sigma}\,q_{t,s}^{k-\rho}\,q_{t,s}^{n-\sigma} \,\delta(s-t)\,ds\,dt
\]

\[
=\frac{1}{2^{n+k}}\,\sum_{\rho=0}^k \,\sum_{\sigma=0}^n 
\binom{k}{\rho}\binom{n}{\sigma}
 (-1)^{\rho} \,\int_{\mathbb{R}^d}\,
\int_{\mathbb{R}^d}\,f(s)\,p_{t,s}^{\rho+\sigma}\,q_{t,s}^{k+n-(\rho+\sigma)} \,\delta(s-t)\,ds\,dt
\]

\[
=\frac{1}{2^{n+k}}\,\sum_{\rho=0}^k \,\sum_{\sigma=0}^n 
\binom{k}{\rho}\binom{n}{\sigma}
 (-1)^{\rho}\, \frac{ \partial^{ \rho+\sigma  } }{ \partial {z}^{\rho+\sigma}}|_{z =0} \,
\int_{\mathbb{R}^d}\,\int_{\mathbb{R}^d}\,f(s)\,
e^{z\,p_{t,s}}\,q_{t,s}^{k+n-(\rho+\sigma)} \,\delta(s-t)\,ds\,dt
\]

\[
=\frac{1}{2^{n+k}}\,\sum_{\rho=0}^k \,\sum_{\sigma=0}^n 
\binom{k}{\rho}\binom{n}{\sigma}
 (-1)^{\rho}\, \frac{ \partial^{ \rho+\sigma  } }{ \partial {z}^{\rho+\sigma}}|_{z =0} \,
\int_{\mathbb{R}^d}\,\int_{\mathbb{R}^d}\,f(s)\,e^{\frac{z}{2}\,p_{t,s}}\,
e^{\frac{z}{2}\,p_{t,s}}\,q_{t,s}^{k+n-(\rho+\sigma)} \,\delta(s-t)\,ds\,dt
\]

\[
=\frac{1}{2^{n+k}}\,\sum_{\rho=0}^k \,\sum_{\sigma=0}^n 
\binom{k}{\rho}\binom{n}{\sigma}
 (-1)^{\rho}\times  
\]

\[
\frac{ \partial^{ \rho+\sigma  } }{ \partial {z}^{\rho+\sigma}}|_{z =0} \int_{\mathbb{R}^d}\,
\int_{\mathbb{R}^d}\,f(s)\,e^{\frac{z}{2}\,p_{t,s}}\,  
\frac{ \partial^{k+n-( \rho+\sigma ) } }{ \partial {w}^{k+n-(\rho+\sigma)}}|_{w =0}  
\left(e^{\frac{z}{2}\,p_{t,s}}\,e^{w\,q_{t,s}}\right) \,\delta(s-t)\,ds\,dt
\]

which,  by Lemma \ref{L}, is equal to

\[
=\frac{1}{2^{n+k}}\,\sum_{\rho=0}^k \,\sum_{\sigma=0}^n 
 \binom{k}{\rho}\binom{n}{\sigma}
 (-1)^{\rho}\times  
\]

\[
\frac{ \partial^{ \rho+\sigma  } }{ \partial {z}^{\rho+\sigma}}|_{z =0} 
\int_{\mathbb{R}^d}\,\int_{\mathbb{R}^d}\,f(s)\,e^{\frac{z}{2}\,p_{t,s}}\,  
\frac{ \partial^{k+n-( \rho+\sigma ) } }{ \partial {w}^{k+n-(\rho+\sigma)}}|_{w =0}  
\left(e^{w\,q_{t,s}}\,e^{\frac{z}{2}\,p_{t,s}}\,e^{w\,z\,\delta(s-t)}\right) \,\delta(s-t)\,ds\,dt
\]

and since,  only the constant term $1$ in the exponential series $ e^{w\,z\,\delta(s-t)} $ will eventually make a nonzero contribution, the above is equal to

\[
=\frac{1}{2^{n+k}}\,\sum_{\rho=0}^k \,\sum_{\sigma=0}^n 
\binom{k}{\rho}\binom{n}{\sigma}
 (-1)^{\rho}\times  
\]

\[
\frac{ \partial^{ \rho+\sigma  } }{ \partial {z}^{\rho+\sigma}}|_{z =0} 
\int_{\mathbb{R}^d}\,\int_{\mathbb{R}^d}\,f(s)\,e^{\frac{z}{2}\,p_{t,s}}\,  
\frac{ \partial^{k+n-( \rho+\sigma ) } }{ \partial {w}^{k+n-(\rho+\sigma)}}|_{w =0}  
\left(e^{w\,q_{t,s}}\,e^{\frac{z}{2}\,p_{t,s}}\right) \,\delta(s-t)\,ds\,dt
\]

\[
=\frac{1}{2^{n+k}}\,\sum_{\rho=0}^k \,\sum_{\sigma=0}^n 
\binom{k}{\rho}\binom{n}{\sigma}
 (-1)^{\rho}\,\frac{ \partial^{ \rho+\sigma  } }{ \partial {z}^{\rho+\sigma}}|_{z =0} 
\int_{\mathbb{R}^d}\,\int_{\mathbb{R}^d}\,f(s)\,
e^{\frac{z}{2}\,p_{t,s}}\, q_{t,s}^{k+n-(\rho+\sigma) }\,e^{\frac{z}{2}\,p_{t,s}} \,\delta(s-t)\,ds\,dt
\]

\[
=\frac{1}{2^{n+k}}\,\sum_{\rho=0}^k \,\sum_{\sigma=0}^n 
\binom{k}{\rho}\binom{n}{\sigma}
 (-1)^{\rho}\,\frac{ \partial^{ \rho+\sigma  } }{ \partial {z}^{\rho+\sigma}}|_{z =0} \,
\int_{\mathbb{R}^d}\,f(t)\,e^{\frac{z}{2}\,p_{t,t}}\, q_{t,t}^{k+n-(\rho+\sigma) }\,
e^{\frac{z}{2}\,p_{t,t}} \,dt
\]

\[
=\sum_{\rho=0}^k \,\sum_{\sigma=0}^n\, 
\binom{k}{\rho}\binom{n}{\sigma}
 \frac{(-1)^{\rho}}{2^{\rho+\sigma}}\,\frac{\partial^{ \rho+\sigma }}{\partial {z}^{\rho+\sigma}}|_{z =0} \,\hat{B}^{k+n+1-(\rho+\sigma) }_{z}(f)
\]

Finally, using (\ref{der}),  (\ref{winfrhpwn}) becomes

\begin{equation}\label{winfrhpwnder}
B^n_k( f)=\sum_{\rho=0}^k \,\sum_{\sigma=0}^n 
\binom{k}{\rho}\binom{n}{\sigma}
 \frac{(-1)^{\rho}}{2^{\rho+\sigma}}\,
\frac{1}{2^{k+n-(\rho+\sigma)}}\,\sum_{m=0}^{k+n-(\rho+\sigma)}\,
\binom{k+n-(\rho+\sigma)}{m}
\end{equation}

\[
\times \sum_{p=0}^{\infty}\,\, (-1)^p\,\binom{ \rho+\sigma }{p}\,B^{ m+p}_{n+k-m-p}(f)
\]

Letting $N:=m+p$, (\ref{winfrhpwnder}) becomes

\begin{equation}\label{winfrhpwnder2}
B^n_k( f)=\sum_{\rho=0}^k \,\sum_{\sigma=0}^n 
\binom{k}{\rho}\binom{n}{\sigma}
 \frac{(-1)^{\rho}}{2^{\rho+\sigma}}\,
\frac{1}{2^{k+n-(\rho+\sigma)}}\,\sum_{m=0}^{k+n-(\rho+\sigma)}\,
\binom{k+n-(\rho+\sigma)}{m}
\end{equation}

\[
\times \sum_{N=m}^{\infty}\,\, (-1)^{N-m}\,\binom{ \rho+\sigma }{N-m}\,B^{ N}_{n+k-N}(f)
\]

Since $n,k$ are fixed, we notice that for each $N$ the coefficient of $B^{ N}_{n+k-N}(f)$ is finite. Thus (\ref{winfrhpwnder2}) is meaninful in the sense of Definition \ref{basic}.

\end{proof}


\begin{thebibliography}{99}

\bibitem{ABIDAQP06} Accardi, L., Boukas, A.:  Renormalized higher powers of white noise (RHPWN) and conformal field theory,  {\em Infinite Dimensional Anal. Quantum  Probab. Related Topics} {\bf 9}, No. 3,  (2006) 353-360. 

\bibitem{ABIJMCS06} \bysame : The emergence of the Virasoro and $w_{\infty}$ Lie algebras through the renormalized higher powers of quantum white noise , {\em International Journal of Mathematics and Computer Science} (2006).

\bibitem{BK91}
Bakas, I.,  Kiritsis,  E.B.: Structure and representations of the $W_{\infty}$ algebra, {\em Prog. Theor. Phys. Supp.} {\bf 102} (1991) 15.

\bibitem{K95}  Ketov, S. V.:{\em Conformal field theory}, World Scientific, 1995.

\bibitem{Z85} Zamolodchikov, A.B.:  Infinite additional symmetries in two-dimensional conformal quantum field theory, {\em Teo. Mat. Fiz.} {\bf 65} (1985), 347--359. 

\end{thebibliography}
\end{document}